\documentclass[12pt,a4paper]{iopart}
\usepackage{graphicx}
\usepackage{color}
\usepackage{iopams}

\newcommand{\ave}[1]{{\langle #1 \rangle}}
\newcommand{\Jcur}{j}

\newcommand{\xket}[1]{{\left\vert #1 \right)}}
\newcommand{\xbra}[1]{{\left( #1 \right\vert}}
\newcommand{\xbraket}[2]{{\left( #1 \vert #2 \right)}}

\begin{document}

\title{Real time evolution at finite temperatures with operator space matrix product states}
\date{November 1, 2013}
\author{Iztok Pi\v{z}orn$^1$, Viktor Eisler$^2$, Sabine Andergassen$^2$, Matthias Troyer$^1$}
\address{
$^1$Theoretische Physik, ETH Zurich, CH-8093 Z\"urich, Switzerland \\
$^2$Faculty of Physics, University of Vienna, Boltzmanngasse 5, A-1090 Wien, Austria}


\begin{abstract}
We propose a method to simulate the real time evolution of one 
dimensional quantum many-body systems at finite temperature by expressing both 
the density matrices and the observables as matrix product states.
This allows the calculation of expectation values and correlation functions as 
scalar products in operator space. 
The simulations of density matrices in inverse temperature and the local 
operators in the Heisenberg picture are independent and result in a grid of 
expectation values for all intermediate temperatures and times. 
Simulations can be performed using real arithmetics with only polynomial growth 
of computational resources in inverse temperature and time for integrable 
systems. 
The method is illustrated for the  XXZ model and the single impurity Anderson 
model. 
\end{abstract}


\section{Introduction}
The simulation of strongly interacting quantum many-body systems is, in 
general, a computationally difficult task. 
Even in the case of one-dimensional quantum systems with finite range 
interactions, where ground state properties have become readily accessible 
using the density matrix renormalization group (DMRG) 
algorithm \cite{whiteprl92}, accessing the dynamics remains hard. 
While, the matrix product states (MPS) \cite{ostlundrommer,vidal,verstraetempo}
that are at the core of DMRG manage to describe the ground states due to their 
weak entanglement \cite{schollwoeckreview,schollwoeckreview2}, the entanglement increases in the 
real time evolution \cite{vidal,schollwoeckreview,tdmrg1,tdmrg2}
making the study of dynamics computationally challenging even for pure states.

Even harder is dynamics at finite temperatures. 
MPS based methods operate on the level of wave functions.
Systems at thermal equilibrium, described by a \emph{mixture} of quantum states, 
thus need to be purified by being described as a pure state in an enlarged 
system, consisting of the system and an auxiliary environment. 
The simulation is then performed on this purified 
system \cite{verstraetempo,FW05} and by tracing over the degrees of freedom of 
the environment, the properties of the original system are obtained.
This idea was first  implemented in Ref.~\cite{verstraetempo,ZV04} for density 
matrices of thermal quantum systems and systems far from equilibrium.
The same concept was used in the simulation of dynamical properties of thermal 
systems where the real time evolution  was computed on the level of the 
purified system and the density operator for the original system was 
obtained by tracing over the environment \cite{barthel1,feiguin}.

A disadvantage of these purification based approaches is that they require 
additional information on the environment, and especially on its dynamics.
In general, these methods are accompanied by an exponential growth of 
computational resources with time and elaborate procedures are required to
control the fast growth of entanglement. 
Recent advances  in the application to the  XXZ model have shown that a smart 
choice  of the environmental dynamics can significantly extend  the 
accessible simulation times \cite{karrasch}. 
However, the question of the \emph{optimal} environment and its dynamics is 
still open. 
Even finding a good choice  of the environmental dynamics requires deep 
knowledge of the system and is far from being straightforward.

Here we propose an alternative strategy, based on operator space concepts and 
matrix product states.
Similarly to the purification approach, we describe the density matrices as 
many-body states; however, our states are not superpositions of quantum states 
but superpositions of operators \cite{prosenznidaricpre,ZV04}. 
In the same way we also describe the observables and their dynamics, known as 
the time dependent DMRG in the Heisenberg picture \cite{prosenznidaricpre,hdmrg}.
An operator space approach was used in \cite{barthel2}, however, still retaining the exponential growth of computational costs. 
The key feature of our approach is the \emph{combination} of both concepts
where the expectation value of an observable at a given time and temperature 
is calculated as a scalar product of two vectors in the operator space, 
one representing the density matrix and the other the real time evolution of 
an operator. 
This way, the simulation of density matrices and the real time evolution of 
operators are \emph{decoupled} and can be performed independently such that only two 
simulations are sufficient to calculate a grid of expectation values at 
various intermediate (inverse) temperatures and times.

The computational costs to simulate the density matrices grow polynomially in 
time due to a low degree of correlation at thermal 
equilibrium \cite{prosenznidaricpizorn}, although such description has mostly 
been used to simulate systems far from the equilibrium \cite{prosenznidaricness}. 
Similarly, the simulation of local operators also exhibits polynomial growth 
of resources \cite{prosenznidaricpre} for integrable systems.
This corresponds to  a logarithmic growth of the operator space equivalent of the entanglement 
entropy (OSEE) \cite{osee}, which is reminiscent of the situation in a pure-state local quantum quench
\cite{ep07,calabrese}.
Our method can therefore be used to simulate the real time evolution of local 
operators in integrable quantum systems in polynomial time.


\section{Methods}
Let us consider a quantum system of $n$ sites with a 
Hilbert space $\mathcal{H}$.  
The expectation value of an operator $a: \mathcal{H} \to \mathcal{H}$ in a 
mixed state described by a density matrix $\rho$, is given by 
$\ave{a} = {\rm tr}(\rho a)$ with ${\rm tr}(\rho)=1$. 
Like the operator $a$, the density matrix is also a linear map 
over $\mathcal{H}$ and these maps form the \emph{operator space} 
algebra $\mathcal{K}$ with an inner product 
$\xbraket{a}{b} = (\textrm{dim}\mathcal{H})^{-1} {\rm tr}(a^\dagger b)$.
The expectation value of $a$ can thus be given as 
\begin{equation}
\ave{a} = \xbraket{\rho}{a} \big/ \xbraket{\rho}{e},
\label{eq:aveA}
\end{equation}
where the bracket notation $\xket{a}$ is used when referring to 
$a \in \mathcal{K}$ and $\xket{e}$ corresponds to the fully mixed state 
(the identity).
In thermal equilibrium for inverse temperature $\beta = (k_{\rm B} T)^{-1}$, 
the expectation values are given by using
$\xket{\rho(\beta)} = \xket{e^{-\beta H}}$ as the density matrix
whereas the time evolution of Eq.~(\ref{eq:aveA}) is obtained by 
replacing $a$ by the Heisenberg operator $a(t) = e^{i t H} a e^{-i t H}$.
Clearly, the simulations for $\rho(\beta)$ and $a(t)$ can be done independently.

The density matrices $\rho(\beta)$ are obtained by the imaginary time 
evolution, starting from  the infinite temperature equilibrium state $\xket{e}$. 
By defining a map \emph{over} the operator space, 
$\hat{\chi}: \mathcal{K} \to \mathcal{K}$, which left-multiplies an operator 
by $H$, i.e. $\hat{\chi} \xket{a} \equiv \xket{H a}$, the density matrix 
$\xket{\rho(\beta)} = \xket{e^{-\beta H}}$ becomes 
\begin{equation}
\xket{\rho(\beta)} = e^{-\beta \hat{\chi} } \xket{e}.
\label{eq:rhobeta}
\end{equation}

Similarly, the Heisenberg evolution $a(t) = e^{i t H}  a e^{-i t H}$ can be 
simulated by introducing a linear map defined as 
$\hat{H} \xket{x} \equiv \xket{ [x, H] }$ for $x \in \mathcal{K}$.
It was shown \cite{osee} that the Heisenberg evolution is hence transformed 
to the Schr\"odinger evolution in the operator space 
\begin{equation}
\xket{a(t)} = e^{-i t \hat{H}} \xket{a(0)}.
\label{eq:at}
\end{equation}
Both $\hat{\chi}$ and $\hat{H}$ have the same form and the same range of interactions as $H$,
if the basis for $\mathcal{K}$ is local (see \ref{app:A}).

Let us now define a suitable basis for $\mathcal{K}$. 
While any basis could in principle be used, it is advantageous to choose a 
basis given by direct products of local basis operators 
$\{ p_{\nu_j}^{[j]} \}$ pertaining to the site $j$, as
$P_{\underline{\nu}} = 
 p_{\nu_1}^{[1]} \otimes p_{\nu_2}^{[2]} \otimes  
    \cdots \otimes p_{\nu_n}^{[n]}$.
These operators are chosen such that they form an orthonormal set. 
A suitable choice for $\{ p_{\nu_j}^{[j]} \}$ depends on the type of 
simulation and a different choice can be made for the 
thermal~(\ref{eq:rhobeta}) and the real time evolution~(\ref{eq:at}). 
For the former, the map $\hat{\chi}$, represented by 
$[ \hat{\chi}]_{\mu,\nu} = 
   (\textrm{dim}\mathcal{H})^{-1} {\rm tr}( P_{\mu}^\dagger \hat{\chi}  P_{\nu} )$,
is real if $\{ P_\nu\}$ are real which in turn 
allows us to simulate the thermal density matrices~(\ref{eq:rhobeta}) with 
real arithmetics.

For the real time evolution~(\ref{eq:at}), where~$\hat{H}$ is represented by 
a matrix 
$[\hat{H}]_{\mu,\nu} = 
 (\textrm{dim}\mathcal{H})^{-1} {\rm tr}( [P_{\mu}^\dagger , P_{\nu}] H)$,
 a different choice for $\{P_{\nu}\}$ is made to maintain the formulation in 
 real arithmetics.
If $P_{\mu}^\dagger = P_{\mu}$, then $[\hat{H}]_{\mu,\nu} = - [\hat{H}]_{\nu,\mu}$ 
and $\hat{H}$ is represented by a hermitian skew-symmetric matrix. 
Together with the imaginary unit in the exponent of~(\ref{eq:at}), the
Heisenberg evolution is generated by a real valued map. 
Finally, due to different operator space bases used in the thermal and the 
Heisenberg evolution, the calculation of expectation values must be preceded 
by a basis transformation, which in our case is a product operator and has 
no significant effects on the computational costs.

For reference we list a few typical operator space bases. 
In the case of spin-$1/2$ models, spinless fermions or hard core bosons the 
basis is given by the products of Pauli matrices 
$p_{\mu_j}^{[j]} = \sigma_j^{\mu_j}$ for $\mu_j \in \{0,1,2,3\}$.
This basis is hermitian and thus suitable for the real time evolution whereas 
the thermal evolution is done using a real local basis 
$\{ \sigma^0, \sigma^1, -i \sigma^2, \sigma^3\}$.
The local basis for spin-$1$ systems is provided by Gell-Mann matrices 
whereas for spin-$3/2$ it can be chosen as 
$\{ S^{(a,b)} = \sigma^a \otimes \sigma^{b}\}$. 
In all cases, the local basis is hermitian and 
its real representation can be found in a straightforward way.

Let us now focus on the MPS ansatz for the operators. 
For simplicity we shall assume a spin-$1/2$ model although the results are 
readily generalized to any many-body quantum system with a local dimension $d$.
The operators $P_{\nu_1,\ldots,\nu_n} = %
    \sigma_1^{\nu_1}\cdots\sigma_{n}^{\nu_n}$ form the 
basis set $\xket{\underline{\nu} } \equiv \xket{P_{\underline{\nu}}}$ and 
 $a \in \mathcal{K}$ is given in terms of a MPS \cite{prosenznidaricpre}
\[
\xket{a} = \sum_{\nu_1,\ldots,\nu_n} 
         {\rm tr}[\mathbf{A}^{[1] \nu_1} \cdots \mathbf{A}^{[n] \nu_n} ]
         \xket{\nu_1,\ldots,\nu_n},
\]
with real matrices $\mathbf{A}^{[j]\nu_j}$ of dimension at most $D\times D$ 
(see e.g. \cite{verstraetempo}) where $D$ is the bond dimension. 
The evolutions~(\ref{eq:rhobeta}) and~(\ref{eq:at}) are simulated using the 
Suzuki-Trotter decomposition and local updates of the MPS, known as the time 
dependent DMRG algorithm \cite{schollwoeckreview,schollwoeckreview2,tdmrg1,tdmrg2} which is 
particularly efficient when $H$ contains at most nearest-neighbor interactions. 

The thermal expectation value $\ave{A(t)}_\beta$ given by~(\ref{eq:aveA}) is
obtained by combining the independent simulations~(\ref{eq:rhobeta}) 
and~(\ref{eq:at}).
Using the same principles, we can simulate $\langle a(t) b\rangle_\beta$ 
where $a(t)$ and $b$ are MPS, by constructing a matrix product operator 
$\hat{B}: x \mapsto b x$.
The result is given by 
$\langle b\, a(t) \rangle_\beta = %
   \xbra{\rho(\beta)} \hat{B} \xket{a(t)} \big/ \xbraket{\rho(\beta)}{e}$ 
which in the MPS language corresponds to calculating an expectation value. 
Similarly $\langle a(t)\, b\rangle_\beta$ is obtained.

\section{Results}
The method proposed here can be used to simulate the real time evolution of a 
wide range of models at finite temperature, for local operators such as the 
spin projection at a given site, a local current, or correlations of local 
operators (e.g. Green's functions). 
We start by analyzing the complexity of both constituents of the method, 
the evolution of the density matrices and the time evolution of operators 
in the Heisenberg picture. 
We quantify the complexity in terms of the OSEE \cite{osee} which is an 
operator space analogue of the entanglement entropy for quantum states, 
defined as $S^\sharp(x) = -{\rm tr}[\hat{R} \log_2\hat{R}]$ 
for $\hat{R} = {\rm tr}_{E}(\xket{x}\xbra{x})$. The bipartition is chosen symmetrically
and thus the environment $E$ in the partial trace includes half of the system.
The effective bond dimension then scales as $\exp(S^\sharp(x))$ with the OSEE.
For density matrices, it was also observed \cite{prosenznidaricpizorn} that the 
OSEE is related
to the quantum mutual information which quantifies the total correlations 
between two parts of the system. 
\begin{figure}
\centering
\includegraphics[width=0.49\columnwidth]{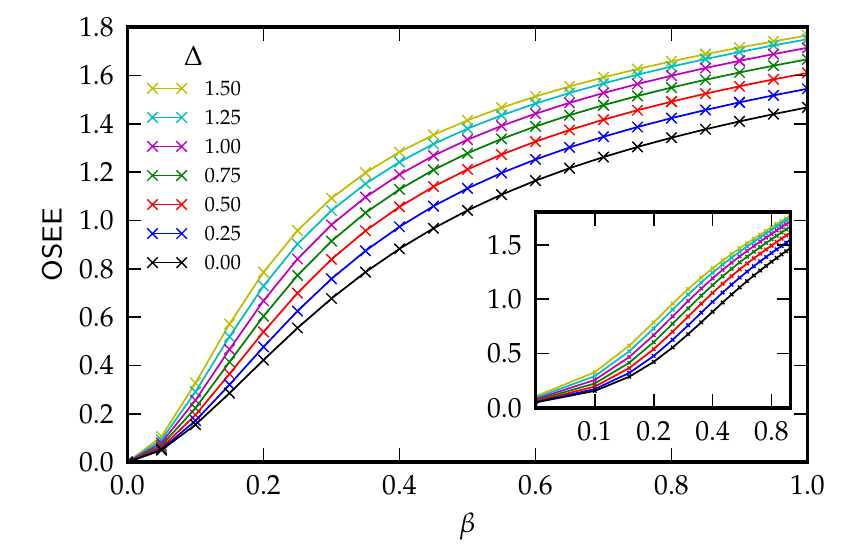}
\includegraphics[width=0.49\columnwidth]{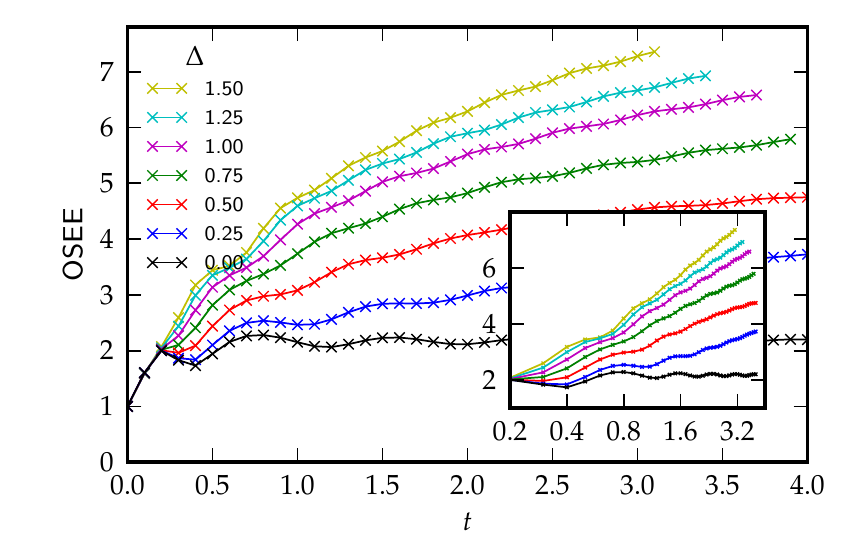}
\caption{
Operator space entanglement entropy (OSEE) in the evolution of density 
matrices $\rho(\beta)$ in the inverse temperature $\beta$ (left) 
and the local current operator $\Jcur_{n/2}(t)$ (right) for the XXZ model 
on $100$ sites with $\Delta$ indicated in the legends. 
Bond dimensions are $D=1024$ and $D=1200$, respectively.
A logarithmic dependence in (imaginary) time is observed in both cases 
(see the insets in a semi-log scale).
}
\label{fig:osee}
\end{figure}

We first consider a  spin-$1/2$ XXZ model with open boundary conditions 
described by the Hamiltonian
\begin{equation}
H = \sum_{l=1}^{n-1} \Big( 
          \sigma_{l}^{x}\sigma_{l+1}^{x} + 
          \sigma_{l}^{y}\sigma_{l+1}^{y} + 
   \Delta \sigma_{l}^{z}\sigma_{l+1}^{z} 
          \Big),
\label{eq:xxz}
\end{equation}
where $\Delta$ is the anisotropy parameter. 
The OSEE of the initial state $\rho(\beta=0)$ is zero 
(the fully mixed state is separable). 
As shown in Fig.~\ref{fig:osee} (left), we observe logarithmic growth of the 
OSEE in the inverse temperature~\cite{prosenznidaricpizorn}.
Furthermore, we simulate the Heisenberg evolution~(\ref{eq:at}) of the 
spin current flowing between sites $m$ and $m+1$,
$\Jcur_m = i (\sigma_m^+ \sigma_{m+1}^- - \sigma_m^- \sigma_{m+1}^{+})$, 
with $\sigma_m^{\pm} = (\sigma_m^x \pm i \sigma_m^y)/2$ for $m = n/2$.
The results in Fig.~\ref{fig:osee} (right) show the logarithmic growth of 
OSEE which is an extension of the results in~\cite{prosenznidaricpre}.
Similar to the evolution of density matrices, we observe an increase
of OSEE for larger anisotropies $\Delta$ where the simulation is more 
expensive.
The logarithmic growth of OSEE corresponds to polynomial growth of the bond 
dimension~\cite{prosenznidaricpre,osee} and the simulation can be performed 
efficiently.
We note that the computational cost is expected to grow 
exponentially \cite{prosenznidaricpre} for non-integrable systems, 
like in other similar methods.

\begin{figure}
\centering
\includegraphics[width=0.6\columnwidth]{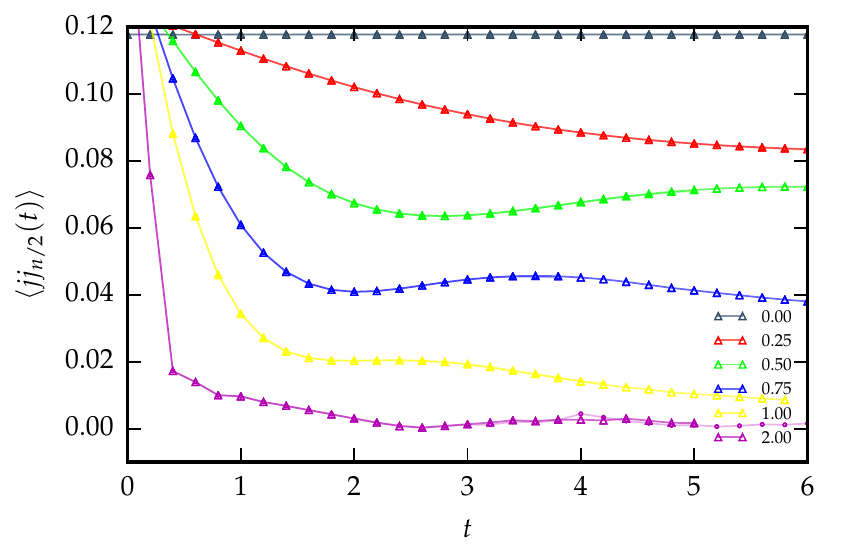}
\caption{
Current correlation function for the  XXZ model on $100$ sites for 
various $\Delta$ and fixed  $\beta=0.25$.
Different symbols correspond to bond dimensions $D=800$
(circles) and $D=1000$ (triangles) for the real time evolution. 
The error in $\rho(\beta)$ is negligible for $D=200$.}
\label{fig:drude}
\end{figure}

The simulation of the density matrices and the local operators can now be 
used to calculate e.g. the spin current correlation function 
$\langle \Jcur \Jcur_{n/2}(t)\rangle_\beta$.
The results are shown in Fig.~\ref{fig:drude}, with an inverse temperature 
$\beta = 0.25$.
The infinite-time limit of these current-correlations yields the Drude weight 
which can be used to signal ballistic transport in various integrable models, 
see \cite{fabianrev}. 
For the XXZ chain similar results were obtained in 
Refs.~\cite{sirker,karrasch,drude}, a detailed discussion is however beyond 
the scope of this manuscript and will be presented elsewhere. 
When compared to Ref. \cite{karrasch} 
(with our time scale multiplied by a factor of 4 due to the spin operators) 
we observe, on one hand, that the reachable times are approximately
twice as large in the gapped regime ($\Delta=2$  in Fig.~\ref{fig:drude})
with a maximal bond dimension $D=1000$.
On the other hand, the computation in the gapless regime, $\Delta<1$, is less
demanding and the curves obtained with a smaller bond dimension $D=800$
are essentially indistinguishable in Fig.~\ref{fig:drude}. This is in complete agreement
with the picture obtained from the OSEE scaling. In turn, for small $\Delta$
the simulation can be pushed beyond the time scale shown in Fig.~\ref{fig:drude},
similarly to Ref. \cite{karrasch}.

\begin{figure}
\centering
\includegraphics[width=0.6\columnwidth]{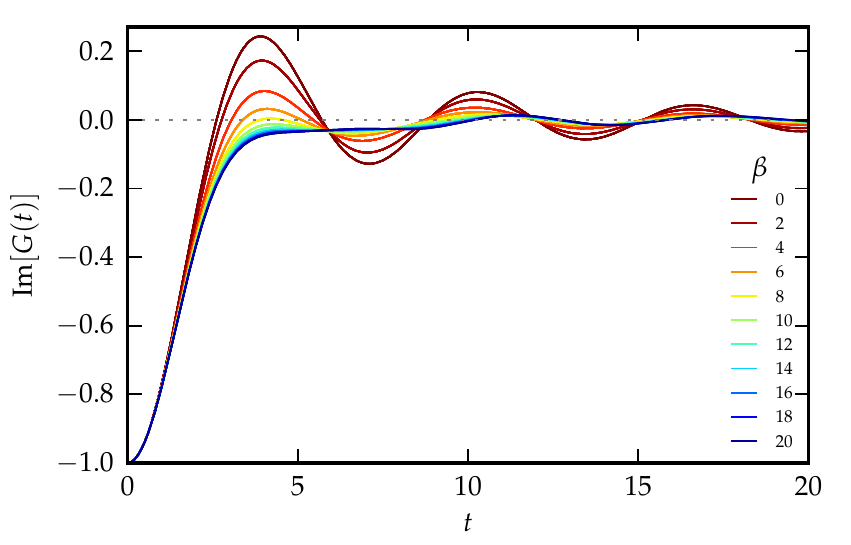}
\caption{Temporal Green's function for the single impurity Anderson
model~(\ref{eq:impuritymodel}) for $U=1$, $\tau_j = \frac{1}{2}$, $\tau_0=0$, 
and the particle-hole symmetric $\epsilon_f=-U/2$ 
where local Coulomb correlations are most pronounced.
}
\label{fig:thermsiam}
\end{figure}

As another illustration of the method we calculate the Green's functions of 
the single impurity Anderson model which can be 
mapped \cite{bulla,weichselbaum} to an XXZ-like model
\begin{equation}
H = \sum_{j=-\infty}^{\infty} 
       \tau_j ( \sigma_j^+ \sigma_{j+1}^- + \sigma_j^- \sigma_{j+1}^+) + 
       U n_{0} n_1 + 
       \epsilon_f (n_{0}+n_{1})
\label{eq:impuritymodel}
\end{equation}
with $n_j = \sigma_j^- \sigma_j^+$ and the interaction strength $U$.
The hopping parameters $\tau_{-1}=\tau_{1}$ are given by the hybridization 
whereas the rest of $\tau_j$ are determined from the discretization of the 
bath (and $\tau_0=0$), see~\cite{weichselbaum} for details.
The interaction term is only present between sites $0$ and $1$ 
(corresponding to spin-$\uparrow$ and spin-$\downarrow$ of the impurity) 
which allows longer acessible times and lower temperatures.
We simulate the single particle Green's function 
$G(t) = -i \langle \{ f_{\uparrow}^\dagger,  f_{\uparrow}(t) \} \rangle_\beta$ 
where $f_{\uparrow}$ is the annihilation operator at the impurity for 
spin-$\uparrow$, written as 
$f_{\uparrow} = (\prod_{j<0}\sigma_{j}^z)\sigma_0^+$ 
due to Jordan-Wigner transformation. 
Despite the nonlocality, the Heisenberg evolution of this operator can be 
simulated efficiently \cite{prosenznidaricpre,osee}. 
We actually simulate Majorana operators 
$w = f_\uparrow + f_{\uparrow}^\dagger$ and 
$w' =  i (f_\uparrow - f_{\uparrow}^\dagger)$ which are hermitian and allow
the simulation in the real arithmetics. 
The results for the imaginary part of the Green's function, 
shown in Fig.~\ref{fig:thermsiam}, can be used to calculate the spectral 
functions (see e.g. Ref.\cite{weichselbaumprl,pruschke,isidori}) by means of 
a Fourier transform which will be in detail presented elsewhere.

\section{Conclusions}
We have proposed a tensor network method to simulate the real time dynamics 
of one dimensional quantum many-body systems at finite temperatures.
The main advantage of the method is that it decouples the real time evolution
from the imaginary evolution in the inverse temperature such that two 
independent simulations give the results for all combinations 
temperature--time.
Furthermore, the simulations are done in real arithmetics.
The method is most useful for integrable systems where the costs grow 
polynomially, and systems close to the integrability where the asymptotic 
exponential growth only appears at later times.
The method can also be used to study nonequilibrium impurity
physics \cite{meisner}, in particular current-voltage characteristics of 
strongly correlated nanostructures in the non-linear response regime.
The proposed framework can be extended to two 
dimensional systems
described in terms of projected entangled pair states \cite{peps} (see
\ref{app:B}); albeit with significantly higher computational costs.

\ack{
We thank F. Verstraete, Lei Wang and C. Karrasch 
for valuable discussions 
and F. Heidrich-Meisner, T. Prosen and M. \v{Z}nidari\v{c} 
for their comments on the manuscript. 
IP acknowledges the collaboration on related projects with 
T. Prosen and M. \v{Z}nidari\v{c}.
SA and VE thank M. Rams and B. Terhal for discussions.
This work was supported by the Swiss National Science Foundation
through the NCCR QSIT, 
by the European Research Council through ERC grants QUERG and SIMCOFE 
and by SFB ViCoM (FWF project ID F4104).
Simulations were run on the Brutus cluster at ETH Zurich 
and on the Vienna Scientific Cluster.
}

\appendix

\section{Algorithm}
\label{app:A}
In Table~\ref{table:algorithm} we present a 
algorithm to simulate the real time evolution at finite temperatures. In particular, we explain how to generate the super operators $\hat{\chi}$ and $\hat{H}$ for a one-dimensional system of $n$ sites with a local dimension $d$ and a Hamiltonian operator $H$. 
Let us first decompose the Hamiltonian into a sum of local product operators
\begin{equation}
H = \sum_\mu H_1^{(\mu)} + \sum_{\nu} H_2^{(\nu)},
\label{eq:ap:H}
\end{equation}
where $H_1^{(\mu)}$ are 
local operators of dimension $d \times d$ and $H_2^{(\nu)}$ are two-site operators of dimension $(d \times d) \times (d \times d)$.
The local Hilbert space for one site is generated by a set of $d^2$ Gell-Mann matrices 
$\{ \sigma^j, \, j=0,1,\ldots,d^2-1\}$ of dimension $d \times d$ where $\sigma^0 = \mathbf{1}$ is the identity. 
The super operators $\hat{\chi}$ and $\hat{H}$ are defined as $\hat{\chi}: a \mapsto H a$ and $\hat{H}: a \mapsto [a,H]$.

\begin{table}
\hspace{-.85cm} \textit{Practical algorithm for the calculation of $\langle b a(t) \rangle_\beta$ }\\[.4cm]
\begin{enumerate}
\item Assume a one-dimensional system of $n$ sites with a local dimension $d$ and Hamiltonian $H$ of the form~(\ref{eq:ap:H})
\item Set up
Gell-Mann matrices $\{ \sigma^j, \, j=0,1,\ldots,d^2-1\}$
\item Construct super-maps $\hat{\chi}$ and $\hat{H}$
\item Construct
a MPS representation of the the unit  $\xket{e}$ 
\item Construct
a MPS representation for the initial operator $\xket{a}$
\item Simulate $\xket{ \rho(\beta) } = {\rm e}^{-\beta \hat{\chi}} \xket{e}$ using tDMRG
and save $\xket{\rho(\beta)}$ for various inverse temperatures $\beta$
\item Simulate $\xket{ a(t) } = {\rm e}^{-i t \hat{H}} \xket{a}$ using tDMRG 
and save $\xket{a(t)}$ for various times $t$
\item Construct
$\hat{b}$ from $b$ such that $\hat{b}: x \mapsto b x$ (in the same way as $\hat{\chi}$ from $H$).
\item For all $t$ and $\beta$, calculate $\xbra{\rho(\beta)} \hat{b} \xket{a(t)}$ and $\xbraket{\rho(\beta)}{e}$.
\end{enumerate}
\caption{Practical algorithm to simulate a time correlation function $\langle b a(t) \rangle_\beta$ at an inverse temperature $\beta$.}
\label{table:algorithm}
\end{table}

The thermal super-map $\hat{\chi}$ thus contains the same number of product operators as the hamiltonian 
\[
\hat{\chi} = \sum_\mu \hat{\chi}_1^{(\mu)} + \sum_\nu \hat{\chi}_2^{(\nu)},
\]
where single-site terms are transformed to $d^2 \times d^2$ matrices 
 \[
 [\hat{\chi}_1^{(\mu)} ]_{j,l} = d^{-1}
 {\rm tr}\big[ (\sigma^j)^\dagger H_1^{(\mu)} \sigma^l \big], \qquad j,l=0,1,\ldots,d^2-1
 \]
and the two-site terms  to $d^2 \times d^2 \times d^2 \times d^2$ tensors
 \[
 [\hat{\chi}_2^{(\nu)} ]_{j_1,j_2,l_1,l_2} = d^{-2}
 {\rm tr}\big[ (\sigma^{j_1} \otimes \sigma^{j_2})^\dagger H_2^{(\nu)} (\sigma^{l_1}\otimes \sigma^{l_2} ) \big].
 \]
 
The super-map $\hat{H}$ contains twice as many local terms as the Hamiltonian operator (due to the commutator), namely
\[
\hat{H} = \sum_{\mu} (\hat{H}_1^{(2\mu-1)} + \hat{H}_1^{(2\mu)}) + 
\sum_{\nu} (\hat{H}_2^{(2\nu-1)} + \hat{H}_2^{(2\nu)}).
\]
The single-site terms are obtained as 
 \[
 [\hat{H}_1^{(2\mu-1)} ]_{j,l} = d^{-1}
 {\rm tr}\big[ (\sigma^j)^\dagger \sigma^l  H^{(\mu)} \big]
 \]
 and
 \[
 [\hat{H}_1^{(2\mu)} ]_{j,l} = d^{-1}
 {\rm tr}\big[ (\sigma^j)^\dagger  H^{(\mu)} \sigma^l  \big].
 \]
 and similarly for the two-site terms.
 
Note that we can and we should (to work in real arithmetics) use a different basis set
$\{ \sigma_\nu \}$ for $\hat{\chi}$ and $\hat{H}$,real matrices for the former and Hermitian for the latter.

The transformation from the Hamiltonian to $\hat{\chi}$ and $\hat{H}$ can thus be treated as a black-box routine, providing the ``Hamiltonian operators'' which can be used in an imaginary (thermal) and a real (Heisenberg picture) time evolution -- using an existing implementation of the time dependent DMRG algorithm (tDMRG) \cite{schollwoeckreview}.

\section{Two-dimensional systems with PEPS}
\label{app:B}
Here we briefly sketch the way how the method can be extended to two dimensional systems.
In principle, the method can already be used for small two-dimensional systems as-is, represented by a matrix product state in a snake-like structure \cite{whiteprl92,schollwoeckreview} which gives remarkably accurate results for the ground states, see e.g. Ref. \cite{depenbrock} for a recent study.
However, for more extended
two-dimensional systems the preferred description is given in terms of Projected Entangled Pair States (PEPS) \cite{peps}. 
Here we show how our method can be used with PEPS. For simplicity we assume a spin-$1/2$ quantum system on a $m \times n$ rectangular lattice.
Any operator can be written as a superposition of basis operators
\[
a = \sum_{\nu_{i,j}} a_{\underline{\nu}} \, \sigma_{1,1}^{\nu_{1,1}}\cdots \sigma_{1,n}^{\nu_{1,n}}\cdots \sigma_{m,n}^{\nu_{m,n}}
\]
for $\nu_{i,j} \in \{0,x,y,z\}$ and the corresponding elements of the operator space $\xket{a}$ can be represented by a PEPS-Ansatz,
\begin{equation}
\xket{a} = \sum_{\nu_{1,1}\cdots \nu_{m,n}} {\rm tr}
\big[ \mathbf{A}^{[1,1] \nu_{1,1}} \cdots \mathbf{A}^{[m,n] \nu_{m,n}} \big] 
\xket{\nu_{1,1},\ldots, \nu_{m,n}}.
\label{eq:peps}
\end{equation}
The trace in~(\ref{eq:peps}) must be understood as a tensor-trace, for details on PEPS see Ref.~\cite{peps}.
The operator $a$ is now represented in terms of rank-$4$ tensors $\mathbf{A}^{[i,j] \nu_{i,j}} \in \mathbb{R}^{D_{\rm l} \times D_{\rm u}  \times D_{\rm d}  \times D_{\rm r}}$ where 
$(D_{\rm l},D_{\rm u},D_{\rm d},D_{\rm r})$, are the dimensions of the bonds connecting the site $[i,j]$ to the neighboring sites (left,up,down,right), respectively. 

The method proposed in the main text can now be used exactly in the same way as in the one-dimensional case, by operating on PEPS instead of MPS. The super-maps $\hat{\chi}$ and $\hat{H}$ are identical to the one-dimensional case and generate the evolution in the Heisenberg picture
\[
\xket{a(t)} = {\rm e}^{- i t \hat{H}} \xket{a}
\]
and the thermal evolution 
of the density matrix
\[
\xket{\rho(\beta)} = {\rm e}^{-\beta \hat{\chi}} \xket{e}.
\]
The only difference to the one-dimensional case is that the computation with PEPS
is significantly more expensive and an accurate description might require 
a considerably increased
computational effort.
In particular,  
the complexity of the real time evolution scales as $O(D^{12})$ (as opposed to $O(D^3)$ for matrix product states). It is not clear at present whether the computational costs still scale polynomially in time like in the one-dimensional case, however, results from the ground state simulations 
are consistent with this conjecture.

\end{document}